\documentstyle[11pt]{article}
\begin{document}
\thispagestyle{empty}
\begin{center}

\vspace{1.8cm} {\bf SYMPLECTIC DEFORMATIONS, NON COMMUTATIVE SCALAR FIELDS AND
FRACTIONAL QUANTUM HALL EFFECT}\\
\vspace{1.5cm} {\bf M. Daoud}$^a${\footnote {\it Facult\'e des
Sciences, D\'epartement de Physique, Agadir, Morocco; email:
m$_-$daoud@hotmail.com}} and {\bf A. Hamama}$^b$
\\\vspace{0.5cm}
$^a$ {\it Max Planck Institute for Physics of Complex Systems,\\ N$\ddot{o}$thnitzer Str. 38, D-01187 Dresden, Germany}\\ \vspace{0.2cm} $^b${\it
High Energy Laboratory, Faculty of Sciences, University Mohamed V,\\
P.O. Box 1014 , Rabat ,
Morocco}\\[1em]

\vspace{3cm} {\bf Abstract}
\end{center}
\baselineskip=18pt
\medskip
We clearly show that the symplectic structures deformations lead,
upon quantization, to quantum theories of non commutative fields.
Two variants of deformations are considered. The quantization is
performed and the modes expansions of the quantum fields are
derived. The Hamiltonians are given and the degeneracies lifting
induced by the deformation is also discussed. As illustration, we
consider the noncommutative chiral boson fields in the context of
fractional quantum Hall effect. A generalized fractional filling
factor is derived and shown to reproduce the Jain Hall states. We
also show that the coupling of left and right edge excitations of
a quantum Hall sample, gives rise a noncommutative chiral boson
theory. The coupling or the non-commutativity induces a shift of
the chiral components velocities. A non linear dispersion relation
is obtained corroborating some recent analytical and numerical
analysis.

\newpage
\section{Introduction}
The noncommutative space time [1-2] has attracted a great deal of
attention during the last decade due to its relevance to quantum
aspects of gravity and its connection with the low energy
description of string theory in the presence of a constant NS-NS
$B$-field [3-4]. Field theories on the Moyal noncommutative space
time have various features which are noticeably different from the
models on the commutative space-time as for instance the UV/IR
mixing [5], twisted symmetries [6-9], twisted statistics [10-15],
etc. Beside the high energy physics, noncommutative field theory
has found applications in condensed matter issues. The most
studied example is the planar system of a collection of fermions
evolving under a strong magnetic field. It is now well established
that the ideas  of noncommutative geometry are relevant in the
context of quantum hall effect in the plane [16] as well as other
spaces with
different geometries and of higher dimensions [17-18].\\
Usually we refer to field theories in noncommutative space-time as
"noncommutative field theories". More recently, a "quantum theory of
noncommutative fields" was introduced [19-25] as a generalization of
noncommutative quantum mechanics [26-27] and it is different from
the quantum field theory constructed over noncommutative space-time.
The theory of noncommutative fields is unrelated to space-time
non-commutativity. The interest on these new kind of quantum field
theories is mainly motivated by the violation of lorentz invariance
and may provide a description of the experimentally observed
matter/antimatter asymmetry.\\

In this work, we shall first be concerned with the quantum theory
of noncommutative scalar fields where the fields and their
conjugate momenta obey deformed equal time commutations rules.
More precisely, unlike the usual case, the quantum fields cease to
commute among themselves. In the first part of this paper we
propose a noncommutative formulation of two dimensional scalar
field theory. This formulation is done from a purely  symplectic
point of view. Indeed, we introduce a deformed symplectic two-form
of the phase space associated with  $(1+1)$-noncommutative field
theory described by two canonical pairs $\varphi^i(x,t)$ and
$\pi^i(x,t)$ $(i = 1, 2)$. Two variants of symplectic deformations
are considered. The first one
$$ \Omega = \sum_i\int dx  \delta \varphi^i\wedge \delta\pi^i +
\frac{1}{2} \sum_{ij}{\cal E}_{ij} \int dx \delta\pi^i\wedge
\delta\pi^j,$$ gives rise upon quantization a theory of
noncommutative fields with equal time commutation relations given
$$[\varphi^i(x,t) , \varphi^j(y,t)] = -i{\cal E}_{ij}\delta(x-y)
{\hskip 1cm}[\pi^i(x,t), \pi^j(y,t)] = 0{\hskip 1cm}[\varphi^i(x,t)
, \pi^j(y,t)] = i\delta^{ij}\delta(x-y)$$ where the
non-commutativity is encoded in the constant parameter ${\cal
E}_{ij} = - {\cal E}_{ji}$. The second kind of symplectic
deformation
$$ \Omega = \sum_i\int dx  \delta\varphi^i \wedge\delta\pi^i -
\frac{1}{2} \sum_{ij}{\cal B}_{ij} \int dx \delta\varphi^i\wedge
\delta\varphi^j,$$ leads to non-commutativity in the momentum
space
$$[\varphi^i(x,t) , \varphi^j(y,t)] = 0
{\hskip 1cm}[\pi^i(x,t), \pi^j(y,t)] = i{\cal
B}_{ij}\delta(x-y){\hskip 1cm}[\varphi^i(x,t) , \pi^j(y,t)] =
i\delta^{ij}\delta(x-y)$$ where the deformation tensor ${\cal
B}_{ij}$ is  antisymmetric and  can be viewed as the dual tensor  of
${\cal E}_{ij}$. In quantizing the theory, we perform a
transformation mapping the deformed symplectic two-form in a
canonical one. It is remarkable that this map turn out to be
identical to the so-called dressing transformation introduced in
[25]. Consequently, the dynamics becomes described by a Hamiltonian
involving terms encoding the deformation effect and are responsible
of the degeneracy lifting of the energy levels of the scalar fields.
It is important to stress that the formulation, presented in this
first part of our work,
start at classical level by deforming the symplectic structure.\\

In other hand, it is well known that planar fermions in a strong
magnetic field are confined in lowest Landau levels and behave
like a rigid droplet of liquid. This is the incompressible quantum
fluid picture proposed by Laughlin [28] which constitutes the
basis of the main advances in this field of research, especially
its connection with noncommutative geometry. Indeed, it was shown
that
 Laughlin states at filling
factor  $1/k$ can be provided by an appropriate noncommutative
finite Chern-Simons matrix model at level $k$ and hence reproduces
the basic features of quantum Hall states [29-30]. In this vain, as
mentioned above the ideas of noncommutative geometry were useful to
study the quantum Hall phenomenon in different geometries and for
arbitrary dimensions [16-18] showing that the effective action for
the edge excitations of a quantum hall droplet is generically given
by a chiral boson action. For instance, in system with boundaries,
like the disc geometry, the excitations reside on the edge of the
droplet and the associated dynamics is described by a (1+1) chiral
boson theory (see
for instance [31]).\\

The main purpose of the second part of this paper is to give a
description of the edge excitations of a quantum Hall droplet
using the noncommutative chiral boson theory. An appropriate
noncommutative action will be defined and shown to reproduce the
basic features of Hall states. The quantization of the proposed
model is performed and shown to imply a fractional filling
involving the the non- commutativity parameter. This provides us
with a model which also reproduces the Jain states [32] for
appropriate values of the deformation parameter. It will be also
pointed out that the action describing the coupling between
excitations of left and right edges on a quantum Hall sample is
equivalent to a  noncommutative chiral action. The coupling
strength between the left and right sectors  play the role of the
deformation parameter. We show that the coupling of excitations
living on right and left edges modify the velocities of the chiral
modes. Finally, we will discuss how the noncommutative chiral
boson fields induces a nonlinear dispersion relation. This result
is corroborated by the field description of quantum Hall edge
reconstruction recently proposed in
[33].\\

The paper is organized as follows. In section 2, a brief review of
the symplectic structure and classical analysis of the massless
scalar field theory is presented. The section 3 concerns the ${\cal
E}$ deformed phase space. We quantize the theory and we give the
mode expansions of deformed scalar fields. Similarly, in section 4,
the ${\cal B}$ deformed phase space is defined. The quantization is
performed and the quantum theory of ${\cal B}$-deformed scalar field
is given. We discuss the lifting degeneracy induced by the
deformation. A general action describing the noncommutative $(1+1)$
chiral boson theory is derived in section 5. The non-commutativity
induces a coupling between left and right chiral sectors. The
obtained action presents many similitude with one introduced in [31]
to classify different hierarchies in abelian fractional Hall effect.
The connection between noncommutative chiral theory and fractional
Hall effect is established in section 6. In fact, using an
appropriate deformed metric, we obtain a generalized fractional
filling factor in term of the deformation tensor strength. We also
discuss the anyonic like statistics of the electron operators and
the deviation from the fermionic statistics caused by the
deformation. In section 7, we write the effective action describing
the coupling between the left and right components of a scalar field. Using the tools
developed in section 5, we show that the coupling induces a shift of
the modes velocities. In section 8, we derive a nonlinear dispersion
relation of noncommutative edge excitations modes of a quantum Hall
droplet. Concluding remarks close this paper.

\section{General considerations}
As we will essentially concerned with deformed phase space of
scalar fields , let us begin by recalling briefly some elements of
symplectic structures of the phase space for (1+1) scalar fields
theory.  More precisely, we introduce  the symplectic two form  to
define the basic geometric structures. This will be useful in
building the phase space of scalar fields with modified symplectic
structures. For this end, We  consider the real massless bosonic
field
\begin{eqnarray}
\lefteqn{ \varphi :  \mathbf{\Sigma}\longrightarrow
\mathbf{R}^2 }\nonumber\\
&& {} ( x , t ) \longrightarrow (\varphi^1( x , t ) ,  \varphi^2(
x , t ))
\end{eqnarray}
described by the action
\begin{eqnarray}
S = \int_{\Sigma}dtdx {\cal L} =\frac{1}{2l}\sum_{i=1,2}
\int_{\Sigma}dtdx((\partial_t\varphi^i)^2 - (\partial_x\varphi^i)^2)
\end{eqnarray}
where the space-time region $\Sigma$ will be taken to be of the form
$[ 0 , l]\times [t_i , t_f]$. The fields are confined in a line
segment of length $l$. To simplify we fix $l = 2\pi$. With the
periodic boundary conditions we can write the fields and their time
derivatives as expansions of a set of modes functions
\begin{eqnarray}
\varphi^i( x , t ) = \frac{1}{\sqrt{ 2\pi}}\sum_{n \in \mathbf{Z}}
q_n^i(t) \exp(-i n x)
\end{eqnarray}
 and
 \begin{eqnarray}
\partial_t\varphi^i( x , t ) = \frac{1}{\sqrt{ 2\pi}}\sum_{n \in \mathbf{Z}} p_n^i(t)
\exp(-i n x).
\end{eqnarray}
The normal or Fourier modes $q_n$ and $p_n$ satisfy the conditions
$ q_{-n} = q_n^{\star}$ and $p_{-n} = p_n^{\star} $ required by
the reality of the fields and  their time
 derivatives. To equip the phase space of the system,
whose variables (coordinates) are the scalar fields
$\varphi^i(x,t)$ and the canonical momentum $\pi^i(x,t) =
\partial_{t}\varphi^i(x,t)$, with a symplectic structure, one
introduce the the canonical one form
\begin{eqnarray}
{\cal A}(t) = \sum_{i=1,2}\int dx
\partial_{t}\varphi^i\delta\varphi^i
\end{eqnarray}
where $\delta $ denotes the exterior derivative on the field space
and the time derivatives of the fields must be treated as
variables. The corresponding  two-form is defined by
\begin{eqnarray}
\Omega_0 = \delta{\cal A}(t) = \sum_{i=1,2}\int dx
\delta\varphi^i\wedge\delta\pi^i.
\end{eqnarray}
We may think $\Omega_0$ as field strength  corresponding to one-form
${\cal A}$ viewed as $U(1)$ gauge potential. It can also be written
as
\begin{eqnarray}
\Omega_0 = \frac{1}{2}\sum_{I,J}\int dx dx'(\Omega_0)_{IJ}(x,x')
\delta\xi^{I}(x)\wedge\delta\xi^{J}(x')
\end{eqnarray}
where we denote the phase space coordinates by $\xi^I(x) \equiv
\xi^{ii'}(x) = \varphi^i(x)$ (resp. $\pi^i(x)$) for $i'=1$ (resp.
for $i'=2$) and $$(\Omega_0)_{IJ}(x,x')\equiv
(\Omega_0)_{IJ}\delta(x-x')=
\delta_{ij}\epsilon_{i'j'}\delta(x-x').$$ It follows that
 the Poisson brackets of two functionals ${\cal F}$ and
${\cal G}$ is given by
\begin{eqnarray}
\{{\cal F} , {\cal G}\} = \sum_{I,J}\int dx
(\Omega_0)^{IJ}\frac{\delta {\cal F}}{\delta\xi^{I}}\frac{\delta
{\cal G}}{\delta\xi^{J}}= \sum_{i=1,2}\int dx \frac{\delta {\cal
F}}{\delta\varphi^{i}}\frac{\delta {\cal G}}{\delta\pi^{i}} -
\frac{\delta {\cal F}}{\delta\pi^{i}}\frac{\delta {\cal
G}}{\delta\varphi^{i}}
\end{eqnarray}
where $\Omega_0^{IJ}$ are the elements of the inverse matrix  of
${\Omega_0}_{IJ}$. Using the equations (3-4) and (6), the
symplectic two-form $\Omega_0$ becomes
\begin{eqnarray}
\Omega_0 =\sum_{i=1,2} \sum_{n\in {\bf Z} }\delta q^i_n \wedge
\delta p^i_{-n}
\end{eqnarray}
and the Poisson brackets take the simple form
\begin{eqnarray}
\{{\cal F} , {\cal G}\} = \sum_{i=1,2}\sum_{n\in {\bf Z}}
\frac{\delta {\cal F}}{\delta q^{i}_n}\frac{\delta {\cal G}}{\delta
p^{i}_{-n}} - \frac{\delta {\cal F}}{\delta p^{i}_{-n}}\frac{\delta
{\cal G}}{\delta q^{i}_n}.
\end{eqnarray}
In particular, the Poisson brackets corresponding to the canonical
coordinates of the phase space $q_n^i$ and $p_n^i$ are given by
\begin{eqnarray}
\{ q_n^i , q_m^j\} = 0 {\hskip 0.5cm} \{ p_n^i , p_m^j\} = 0 {\hskip
0.5cm} \{  q_n^i, p_m^j\} = \delta_{i,j}\delta_{m+n,0}.
\end{eqnarray}
The Hamiltonian given by
\begin{eqnarray}
H_0 = \sum_{i=1,2}\int dx \pi^i\partial_t\varphi^i - \int dx {\cal
L} = \frac{1}{2}\sum_{in} (p_n^i p_{-n}^i + n^2 q_n^iq_{-n}^i).
\end{eqnarray}
is the generator of time translation. The time evolution of a
function ${\cal F}$ is given by the Hamilton's equation of motion
$\dot{{\cal F}}=\{ {\cal F} , H_0\}$ that gives
\begin{eqnarray}
\frac{dq_n^i}{dt} = \{ q_n^i, H_0\} =  p_{n}^i {\hskip 1cm}
\frac{dp_{n}^i}{dt} = \{ p_{n}^i, H_0\} =-n^2q_{-n}^i.
\end{eqnarray}
To pass over the quantum theory in the Heisenberg picture, all
canonical variables become Heisenberg operators satisfying
commutation relations corresponding to Poisson brackets as
\begin{center}
(Poisson Bracket) $\rightarrow$ $-i$ (commutator).
\end{center}
The equations of motion (13) can be obtained directly from
Euler-Lagrange equations. The symplectic procedure discussed in
this section has the advantage, as it will be clear below, to be
more adequate to deal with modified or deformed symplectic
structures.

\section{ ${\cal E}$-Deformed scalar fields}

\subsection{${\cal E}$-Deformed phase space}
In this section, we consider two abelian fields $\varphi^1(x,t)$
and $\varphi^2(x,t)$ described by the Hamiltonian (12) and we
suggest to replace the usual symplectic two-form $\Omega_0$ by
\begin{eqnarray}
\Omega = \Omega_0 + \frac{1}{2} {\cal E}_{ij} \int dx
\delta\xi^{i2}(x) \wedge\delta\xi^{j2}(x).
\end{eqnarray}
The new closed two-form $\Omega$ rewrites in a compact form as
\begin{eqnarray}
\Omega = \frac{1}{2}  \int dx \Omega_{IJ} \delta\xi^{I}(x)
\wedge\delta\xi^{J}(x).
\end{eqnarray}
The constant antisymmetric tensor ${\cal E}_{ij} = \theta
\epsilon_{ij}$ encodes the modification or the deformation of the
symplectic structure providing the non trivial metric
\begin{eqnarray}
\Omega_{IJ} = \delta_{ij}\epsilon_{i'j'} + {\cal
E}_{ij}\delta_{i'2}\delta_{j'2}
\end{eqnarray}
which is non degenerate (det$\Omega \neq 0$).  In order to establish
the connection between the classical and quantum theory, we
introduce the Poisson brackets defined by
\begin{eqnarray}
\{{\cal F} , {\cal G}\} = \sum_{I,J}\int dx \Omega^{IJ}\frac{\delta
{\cal F}}{\delta\xi^{I}}\frac{\delta {\cal G}}{\delta\xi^{J}}
\end{eqnarray}
where $\Omega^{IJ}$ is the inverse matrix of $\Omega_{IJ}$ (16).
Using the equations (3-4), the symplectic form $\Omega$ and the
associated Poisson brackets rewrite
\begin{eqnarray}
\Omega = \delta q^{i}_{n}\wedge\delta p^{i}_{-n}+\frac{1}{2}{\cal
E}_{ij}\delta p^{i}_{n}\wedge\delta p^{j}_{-n}
\end{eqnarray}
and
\begin{eqnarray}
\{{\cal F} , {\cal G} \} = \sum_{in} \frac{\delta {\cal
F}}{\delta q^i_m}\frac{\delta {\cal G}}{\delta p^i_{-n}} -
\frac{\delta {\cal F}}{\delta p^i_{-n}} \frac{\delta {\cal
G}}{\delta q^i_n} - \sum_{ijn} {\cal E}_{ij}\frac{\delta {\cal
F}}{\delta q^i_n} \frac{\delta {\cal G}}{\delta q^j_{-n}}
\end{eqnarray}
respectively, in terms of the normals modes and their conjugate
momenta. The Hamiltonian vector fields $X_{\cal F}$, associated to the functional
${\cal F}$, is given by

$$X_{\cal F} = \sum_{in} X^i_n \frac{\delta
}{\delta q^i_n} + Y^i_n\frac{\delta }{\delta p^i_{-n}}$$ where
$$ X^i_n= \frac{\delta {\cal F}}{\delta p^i_{-n}} - {\cal E}_{ij} \frac{\delta{\cal F}
}{\delta q^j_{-n}} {\hskip 2cm} Y^i_{n} = -\frac{\delta {\cal F}
}{\delta q^i_{n}}$$ such that the interior contraction of $\Omega$
with $X_{\cal F}$ verifies $i(X_{\cal F})\Omega = \delta{\cal F}$.
In particular, from the equation (19), one gets the Poisson
brackets of the phase space variables
\begin{eqnarray}
\{q^i_n , q^j_n \} = -{\cal E}_{ij}\delta_{m+n,0} {\hskip 1cm}
\{p^i_m , p^j_n \} = 0 {\hskip 1cm} \{q^i_n , p^j_{-m} \} =
\delta_{m,n}\delta_{i,j}
\end{eqnarray}
reflecting a deviation from the canonical brackets. At this stage,
it is remarkable that the symplectic form and the Poisson brackets
can be converted in the canonical ones by mean of the following
transformation
\begin{eqnarray}
Q^i_n = q^i_n - \frac{1}{2} {\cal E}^{ik} p^k_{n} {\hskip 1.5cm }
P^i_n = p^i_{n}.
\end{eqnarray}
with summation over  $k$. Indeed, one can check that
\begin{eqnarray}
\{Q^i_n , Q^j_m\} = 0 {\hskip 1cm}\{P^i_n , P^j_m\} = 0  {\hskip
1cm}\{Q^i_n , P^j_m\} = \delta_{m+n,0}\delta_{ij},
\end{eqnarray}
and that the symplectic two-form (18) becomes
\begin{eqnarray}
\Omega = \sum_{in} dQ^i_n \wedge dP^i_{-n}.
\end{eqnarray}
The Hamiltonian (12)  is now given by
\begin{eqnarray}
H = \frac{1}{2}\bigg[ \sum_{in} \frac{1}{4}(4+\theta^2n^2)
P^i_nP^i_{-n} + n^2Q^i_nQ^i_{-n} + n^2\theta
\sum_j\epsilon^{ij}P^i_{-n}Q^j_n\bigg]
\end{eqnarray}
 in terms of the new
dynamical modes of the theory. Evidently the $\theta$-dependent
term in $H$ arises from the deformation of the symplectic
structure. This suggests that the fields $\varphi^1$ and
$\varphi^2$ interacts with a certain internal potential $ V_{int}
$
 to have $H = H_0 + V_{int}$ where $H_0$ is the free hamiltonian
given by (12) modulo the substitution $(q,p) \rightarrow (Q,P) $.
Then, the interacting potential $V_{int}$  induces a modification
of the symplectic structure and a deformation of the phase space
geometry. Further, it is clear that the classical scalar field
theory described by the modified symplectic two-form (14) and the
free Hamiltonian (12) is equivalent to the description given by
the canonical symplectic form $\Omega$ (23) and the Hamiltonian
$H$ (24) expressed in terms of the new phase space variables
($Q^i_n$ , $P^i_n$) and involving
interacting harmonic modes.\\

\subsection{Quantization}
Upon replacing the variables by operators with commutation rules
given by $i$ times the Poisson brackets ( $\{  {\hskip 0.1cm}  ,
{\hskip 0.1cm} \} \longrightarrow -i   [ {\hskip 0.1cm}  ,  {\hskip
0.1cm} ]$), we get
\begin{eqnarray}
[Q^i_n , Q^j_m] = 0 {\hskip 1cm}[P^i_n , P^j_m]= 0 {\hskip
1cm}[Q^i_n , P^j_{-m}]= i\delta_{m+n,0}\delta_{i,j}
\end{eqnarray}
 It follows that the
operators $q^i_n $ and $p^i_n$, corresponding to the
dynamical variables of the classical theory, satisfy the following
commutation relations
\begin{eqnarray}
[q^i_n , q^j_m]= -i{\cal E}_{ij}\delta_{m+n,0}{\hskip 1cm} [p^i_n ,
p^j_m] = 0 {\hskip 1cm} [q^i_n , p^j_{-m}]=
i\delta_{i,j}\delta_{n,m},
\end{eqnarray}
 and the equal time
commutation relations are given by
\begin{eqnarray}
[\varphi^i(x,t) , \varphi^j(y,t) ] = -i {\cal E}^{ij}  \delta(x-y)
\end{eqnarray}
\begin{eqnarray}
 [ \pi^i(x,t) , \pi^j(y,t)] = 0
\end{eqnarray}
\begin{eqnarray}
[ \varphi^i(x,t), \pi^j(y,t)] = i\delta_{ij}\delta(x-y).
\end{eqnarray}
Introducing the operators
\begin{eqnarray}
a^i_n = \sqrt{\frac{\Delta_n}{2}} (Q^i_n + i
\frac{P^i_{n}}{\Delta_n}) {\hskip 1cm} a^{i+}_n =
\sqrt{\frac{\Delta_n}{2}} (Q^i_{-n} - i \frac{P^i_{-n}}{\Delta_n})
\end{eqnarray}
with $\Delta_n = 2|n|(4+\theta^2n^2)^{-\frac{1}{2}}$, the
Hamiltonian $H$ can be also written as
\begin{eqnarray}
 H = H_{n=0}  + \frac{1}{2}\sum_{in\neq
0}\bigg[\frac{|n|}{2}\sqrt{4+\theta^2n^2}(a^i_na^{i+}_n +
a^{i+}_{-n}a^i_{-n}) + i \sum_j \theta n^2 \epsilon^{ij}
a^{i+}_na^j_n\bigg]
\end{eqnarray}
where
\begin{eqnarray}
H_{n=0} = \frac{1}{2}[(P^1_0)^2 + (P^2_0)^2 ]
\end{eqnarray}
stands for the zero mode energy contribution.  The diagonalization of
the Hamiltonian
$H-H_{n=0}$ can be performed by means of the new Weyl
operators
\begin{eqnarray}
A^1_n = \frac{1}{\sqrt{2}}(a^1_n -i a^2_n){\hskip 0.5cm}A^{1+}_n =
\frac{1}{\sqrt{2}}(a^{1+}_n +i a^{2+}_n){\hskip 0.5cm}A^2_n =
\frac{1}{\sqrt{2}}(a^1_n +i a^2_n){\hskip 0.5cm}A^{2+}_n =
\frac{1}{\sqrt{2}}(a^{1+}_n -i a^{2+}_n).
\end{eqnarray}
Substituting (33) in (31), one obtain
\begin{eqnarray}
H-H_{n=0} = \sum_{n\neq 0} (\omega_n^- A^{1+}_nA^1_n + \omega_n^+
A^{2+}_nA^2_n)
\end{eqnarray}
where
\begin{eqnarray}
\omega^{\pm}_n = \frac{|n|}{2}\sqrt{4+\theta^2n^2} \pm \theta n^2
\end{eqnarray}
It is easy seen that the deformation induces a lifting of the
degeneracies of the spectrum. Note that the modes frequencies
satisfy the relations
$\omega^{\pm}_n = \omega^{\pm}_{-n}$ and $\omega^+_n (\theta) =
\omega^-_n(-\theta)$.\\
The dynamics of this system is described by the Heisenberg equations
given by
\begin{eqnarray}
\frac{dA^1_n}{dt} = -i[ A^1_n , H ] = -i \omega_n^- A_n^1 {\hskip
1cm}\frac{dA^2_n}{dt} = -i[ A^2_n , H ] = -i \omega_n^+ A_n^2,
\end{eqnarray}
and the corresponding solutions are
\begin{eqnarray}
A^1_n(t) = \hat{A}_n^1\exp(-i \omega_n^-t)  {\hskip 1cm}A^1_n(t) =
\hat{A}_n^2\exp(-i \omega_n^+t)
\end{eqnarray}
where the operators $\hat{A}^1_n$ and $\hat{A}^2_n$ are
time-independent. Consequently, using the equations (21), (30) and
(33), we obtain the normal modes as
\begin{eqnarray}
q^1_n(t) = \frac{1}{2}\bigg[\Lambda_n^+ \bigg(
\hat{A}^1_{n}\exp(-i\omega_n^-t) +
\hat{A}^{1+}_{-n}\exp(+i\omega^-_nt)\bigg) + \Lambda_n^- \bigg(
\hat{A}^2_{n}\exp(-i\omega_n^+t) +
\hat{A}^{2+}_{-n}\exp(+i\omega_n^+t)\bigg)\bigg]
\end{eqnarray}
 and
\begin{eqnarray}
q^2_n(t) = \frac{i}{2}\bigg[\Lambda_n^+ \bigg(
\hat{A}^1_{n}\exp(-i\omega_n^-t) -
\hat{A}^{1+}_{-n}\exp(+i\omega^-_nt)\bigg) + \Lambda_n^- \bigg(
\hat{A}^{2+}_{-n}\exp(+i\omega_n^+t)-\hat{A}^2_{n}\exp(-i\omega_n^+t)\bigg)\bigg
]
\end{eqnarray}
where the $\theta$-dependent constants $\Lambda_n^{\pm}$ are defined
by
\begin{eqnarray}
\Lambda_n^{\pm} = \frac{1}{\sqrt{\Delta_n}} \pm
\frac{\theta}{2}\sqrt{\Delta_n}.
\end{eqnarray}
It is clear from equation (34) that the non-commutativity induces
a lifting of the energy levels degeneracies. This feature is very
similar to the Landau problem in quantum mechanics. The
Hamiltonian (34) is given by a sum of two independents one
dimensional harmonic oscillators. For $\theta = 0$, we have
$\omega_n^+ = \omega_n^-$ and the Hamiltonian becomes a
superposition of multi-modes two dimensional harmonic oscillators.

\section{${\cal B}$- Deformed scalar fields }

\subsection{${\cal B}$-Deformed phase space }

Hereafter we use the symbols $\Omega , \theta , P  , Q, a$ and $A$
that do not should be confused with ones
introduced in the previous section. To begin we consider the closed
non-degenerate symplectic two-form
\begin{eqnarray}
\Omega = \Omega_0 - \theta\int dx
\delta\varphi^1(x)\wedge\delta\varphi^2(x)
\end{eqnarray}
where the added term involves only the fields $\varphi^1$ and
$\varphi^2$. It can also be writing as
\begin{eqnarray}
\Omega = \Omega_0 - \frac{1}{2}{\cal B}_{ij}\int dx
\delta\xi^{i1}(x)\wedge\delta\xi^{j1}(x)
\end{eqnarray}
where the constant antisymmetric tensor ${\cal B}_{ij}$ is defined
by
$${\cal B}_{ij} = \theta\epsilon_{ij}$$
in term of the non-commutativity parameter $\theta$. As discussed in
the section 2, It is more appropriate to rewrite $\Omega$ in the
following compact form
\begin{eqnarray}
\Omega = \frac{1}{2}\sum_{IJ}\int
dxdx'\Omega_{IJ}(x,x')\delta\xi^{I}(x)\wedge\delta\xi^{J}(x')
\end{eqnarray}
where
\begin{eqnarray}
\Omega_{IJ}(x,x')= \Omega_{IJ}\delta(x-x')=
(\delta_{ij}\epsilon_{i'j'} - \delta_{i'1}\delta_{j'1}{\cal
B}_{ij})\delta(x-x').
\end{eqnarray}
Using the modes expansions given by equations (3) and (4), the
modified symplectic two-form can be expressed as
\begin{eqnarray}
\Omega = \sum_{in} \delta q^i_n \wedge \delta p^i_{-n} -
\frac{1}{2}\sum_{ijn} {\cal B}_{ij} \delta q^i_n \wedge \delta
q^j_{-n}.
\end{eqnarray}
Using the definition (17), the Poisson brackets reads
\begin{eqnarray}
\{{\cal F} , {\cal G} \}  = \sum_{in} \frac{\delta {\cal F}}{\delta
q^i_n}\frac{\delta {\cal G}}{\delta p^i_{-n}} - \frac{\delta
f}{\delta p^i_{-n}} \frac{\delta {\cal G}}{\delta q^i_n} +
\sum_{ijn} {\cal B}_{ij}\frac{\delta {\cal F}}{\delta p^i_n}
\frac{\delta {\cal G}}{\delta p^j_{-n}},
\end{eqnarray}
and in particular the fundamental Poisson brackets are given by
\begin{eqnarray}
\{ q^i_n, q^j_m\} = 0 {\hskip 1cm}\{q^i_n , p^j_{m} \} =
\delta_{ij}\delta_{m+n,0} {\hskip 1cm}\{p^i_n , p^j_m \} = {\cal
B}_{ij}\delta_{m+n,0}.
\end{eqnarray}
Contrary to the previous section, the Poisson brackets between the
variables $q^i_n$ give zero and the canonical momentum variables
acquire non-vanishing Poisson brackets and upon quantization they
become non-commuting operators. This reflects the non-commutativity
in the momentum space induced by the deformation of the canonical
symplectic two-form $\Omega_0$ (see equation (41)).\\
To achieve the physical implications of the modification of the
symplectic two-form $\Omega_0$, we follow a similar procedure that
one used in the previous section. In this sense, we start by
converting $\Omega$  in a canonical two-form
 to perform the quantization of the model. Indeed, the
modified symplectic two-form $\Omega$ can be expressed as
\begin{eqnarray}
\Omega =  \sum_{in} \delta Q^i_n\wedge \delta P^i_{-n}
\end{eqnarray}
in terms of the new dynamical variables $Q^i_n$ and $P^i_n$ defined
as follows
\begin{eqnarray}
Q^i_n = q^i_n  {\hskip 1cm}P^i_n = p^i_n - \frac{1}{2} {\cal
B}_{ij}q^j_{n}
\end{eqnarray}
with a summation over repeated indices. The Poisson bracket (46)
becomes canonical and in particular, we have the canonical
relations
\begin{eqnarray}
\{Q^i_n , Q^j_m\} = 0 {\hskip 1cm}\{P^i_n , P^j_m\} = 0  {\hskip
1cm}\{Q^i_n , P^j_m\} = \delta_{m+n,0}\delta_{ij}.
\end{eqnarray}
Using (49), the Hamiltonian (12) reads
\begin{eqnarray}
H =\frac{1}{2}\sum_{in}\bigg[   P^i_n P^i_{-n}+
  \frac{1}{4}(4n^2 + \theta^2) Q^i_n Q^i_{-n} +\theta
\sum_{j} \epsilon^{ij}P^i_nQ^j_{-n} \bigg].
\end{eqnarray}
At this level, we have derived the necessary tools needed
to quantize the model under consideration.

\subsection{Quantization}

The quantization analysis now, though identical in spirit to the
previous one, differs in detail. The correspondence principle leads
to the commutation rules
\begin{eqnarray}
[Q^i_n , Q^j_m] = 0 {\hskip 1cm}[P^i_n , P^j_m] = 0  {\hskip 1cm}
[Q^i_n , P^j_m] = i \delta_{ij}\delta_{m+n,0}.
\end{eqnarray}
Consequently, we have the equal time commutation relations
\begin{eqnarray}
[\varphi^i(x,t) , \varphi^i(y,t)] = 0 {\hskip 1cm} [\pi^i(x,t),
\pi^j(y,t)] = i{\cal B}_{ij}\delta(x-y) {\hskip 1cm}
[\varphi^i(x,t) , \pi^j(y,t)] = i\delta_{ij}\delta(x-y).
\end{eqnarray}
Passing in the Schwinger representation, the quantized hamiltonian
$H$ becomes
\begin{eqnarray}
H = \sum_{in} \frac{1}{2} \omega_n (a^i_n a^{i+}_n + a^{i+}_na^i_n)
+ i \theta \sum_{nij}\epsilon_{ij}a^{i+}_na^j_n
\end{eqnarray}
in terms of the creation and annihilation operators defined by
\begin{eqnarray}
a^i_n = \sqrt{\frac{\omega_n}{2}}\bigg( Q^i_n + i
\frac{P^i_{n}}{\omega_n}\bigg) {\hskip 1cm}
 a^{i+}_n = \sqrt{\frac{\omega_n}{2}}\bigg( Q^i_{-n} - i
\frac{P^i_{-n}}{\omega_n}\bigg)
\end{eqnarray}
where $2\omega_n =\sqrt{4n^2 + \theta^2} $. To remove the angular
momentum contribution from $H$ (the last term in (54)), we use the
Weyl operators defined by (33) and we get
\begin{eqnarray}
H = \sum_{n}  \omega_n^- A^{1+}_nA^1_n  + \omega_n^+A^{2+}_nA^2_n
\end{eqnarray}
where $\omega_n^{\pm} = \omega_n \pm \theta/2$. Solving the
Heisenberg equations of motion satisfied by the operators $A^i_n$ (
$i = 1, 2$), one can see the normal modes of the quantum fields
$\varphi^1(x,t)$ and $\varphi^2(x,t)$ are given by
\begin{eqnarray}
q^1_n(t) = \frac{1}{2\sqrt{\omega_n}} \bigg(
\hat{A}^1_{n}\exp(-i\omega_n^-t) +
\hat{A}^{1+}_{-n}\exp(+i\omega^-_nt)+
\hat{A}^2_{n}\exp(-i\omega_n^+t) +
\hat{A}^{2+}_{-n}\exp(+i\omega_n^+t)\bigg)
\end{eqnarray}
and
\begin{eqnarray}
q^2_n(t) = \frac{i}{2\sqrt{\omega_n}} \bigg(
\hat{A}^1_{n}\exp(-i\omega_n^-t) -
\hat{A}^{1+}_{-n}\exp(+i\omega^-_nt)
+\hat{A}^{2+}_{-n}\exp(+i\omega_n^+t)-\hat{A}^2_{n}\exp(-i\omega_n^+t)\bigg)
\end{eqnarray}
Here again the operators $\hat{A}^i_n$ are time independents. It
is easily seen from (56)
 that due to the phase space non-commutativity,
the energy levels are not degenerates. Thus, it seems that the deformation induces
a coupling between the fields which removes the degeneracies of
the energy levels.

\section{Noncommutative chiral field space }
\subsection{Commutative chiral field space }
This section  is devoted to chiral boson theory in
connection with quantum Hall effect. In this respect, let us give a brief review of some
basic results in order to  explain our purpose.
Consider a Quantum Hall state on a disc (of unit radius)  with
filling factor $\nu$.  Following the Laughlin picture, we known that the edge excitations might have
many branches $I = 1, 2, \cdots, N$ residing on the edge of the
quantum droplet. It is also now well established that the edge
dynamics, associated with disc geometry, can be described by a
chiral boson field compactified on a circle. Thus, let us
consider a massless scalar field $\phi$ in (1+1) Minkowski
space-time where the metric has the diagonal elements $(1 , -1)$.
The field
$$ \phi = \phi_+ + \phi_-$$
is a superposition of the left $(\phi_- \equiv \phi(x-t))$ and right
$(\phi_+ \equiv \phi(x+t))$ moving components which are known as
chiral bosons. We define the partial derivatives $\partial_0 $ and  $\partial_1 $ to refer to
 differentiations with $t$ and $x$. The action describing the
dynamics of left or right chiral boson is given by
\begin{eqnarray}
S_{s} = \int dxdt [s \partial_1\phi_{s}\partial_0\phi_{s} -
(\partial_1\phi_{s})^2]
\end{eqnarray}
where $s = +$ or $-$ respectively and the velocity is normalized to unity. The corresponding Hamiltonians
are
$$ H_{s} = \int dx (\partial_1\phi_{s})^2.$$
Since the fields live on a compact space, they must satisfy the
boundary condition
$$\phi_{s}(2\pi,t) -  \phi_{s}(0,t)= 2s\pi. $$
The general solutions of the equations of motion arising from (59)
and compatible with the boundary conditions are
\begin{eqnarray}
\phi_{s}(x,t) = \alpha^s_0 + \bar{\alpha}^s_0(t+sx) + i \sum_{n\neq
0}\frac{\alpha^s_n}{n} e^{-in(t+sx)}.
\end{eqnarray}
Upon quantization the field $\phi_s$ and its corresponding canonical
momenta $\pi_s$ (its space derivative) satisfy the commutation rules
\begin{eqnarray}
[\phi_s(x) , \phi_{s'}(x')] = -is\delta_{ss'}\varepsilon(x-x')
\end{eqnarray}
\begin{eqnarray}
[\phi_s(x) , \pi_{s'}(x')] = i\delta_{ss'}\delta(x-x'),
\end{eqnarray}
\begin{eqnarray}
[\pi_s(x),\pi_{s'}(x')] =is\delta_{ss'}\delta'(x-x')
\end{eqnarray}
where $\varepsilon(x-y)$ represents the Heaviside function and
$\delta'(x-y)$ denote the spatial derivative of the delta
function. The coefficients $\alpha^s_0$, $\bar{\alpha}^s_n$ and
$\alpha^s_n$ of the expansion (60) become operators satisfying the
commutations relations
\begin{eqnarray}
[\alpha^s_0,\bar{\alpha}^{s'}_0] =i\delta_{ss'} {\hskip 0.5cm}
[\alpha^s_n , \alpha^{s'}_m] =n\delta_{ss'}\delta_{n+m}{\hskip
0.5cm} others = 0
\end{eqnarray}
acting on a bosonic Fock space, whose vacuum $\vert 0 \rangle$ is
defined by
$$\alpha_n^s \vert 0 \rangle = 0.$$

\subsection{Noncommutative space of fields }

To obtain a quantum theory of noncommutative chiral fields
describing the edge excitations having many branches $(I = 1, 2,
\cdots, N)$ , we start by considering $N$ left moving chiral
fields $\Phi_I$ travelling, on the edge of an incompressible Hall
droplet, with velocities normalized to unity. They satisfy the
commutations rules
\begin{eqnarray}
[\Phi_I(x) , \Phi_{J}(x')] = -i\delta_{IJ}\varepsilon(x-x')
\end{eqnarray}
and are described by the action
\begin{eqnarray}
S = \sum_{I} \int dtdx (\partial_0\Phi_I) (\partial_1\Phi_I) -
(\partial_1\Phi_I)(\partial_1\Phi_I).
\end{eqnarray}
We suggest to replace the commutation relation (65) by
\begin{eqnarray}
[\Phi_I(x) , \Phi_{J}(x')] =
-i\Omega_{IJ}(\theta)\varepsilon(x-x').
\end{eqnarray}
This corresponds to a deformation of the symplectic structure of
the fields space. From a physical point of view, this deformation
can be viewed as a sort of coupling between the $N$ quantum Hall
branches. The parameter $\theta$  encodes the non-commutativity of
the fields space and the matrix $\Omega$ reduces to unit matrix
when the deformation vanishes ($\Omega_{IJ} \to \delta_{IJ}$). The
matrix $\Omega$ in (67) is real, symmetric and it is assumed to be
invertible. We assume also that the dynamics is governed by the
the Hamiltonian
\begin{eqnarray}
H = \sum_I \int dx :(\partial_1\Phi_I)^2:
\end{eqnarray}
where the symbol :: denotes the normal ordering. The Heisenberg
equation
\begin{eqnarray}
-i\frac{d\Phi_I}{dt} = [ H , \Phi_I]
\end{eqnarray}
gives
\begin{eqnarray}
\partial_0\Phi_I = \sum_J \Omega_{IJ}\partial_1\Phi_J
\end{eqnarray}
which  rewrites also as
\begin{eqnarray}
\partial_0^2\Phi_I = \sum_J (\Omega^2)_{IJ}\partial_1^2\Phi_J.
\end{eqnarray}
The last equation is easily solved  by mean of a unitary matrix $U$ which
diagonalizes the matrix $\Omega^2$. Indeed, setting $$\Psi_I = \sum_J
U_{IJ}\Phi_J,$$ one has
\begin{eqnarray}
\partial_0^2\Psi_I = \lambda_I \partial_1^2\Psi_I
\end{eqnarray}
where $\lambda_I$ are the eigenvalues of $\Omega^2$. Consequently,
it clear that the velocities of the fields $\Phi_I$ are given by
\begin{eqnarray}
v_I = \pm \sqrt{\lambda_I}
\end{eqnarray}
where the positive $v_I$ correspond to a left moving branch and
negative value to a right moving one.  It is easily seen from the
equations (72) and (73) that the velocities of the chiral fields
are modified. This is the effect of the deformation of symplectic
structure. This result agrees with one recently obtained in [19].
Obviously, the eigenvalues $\lambda_I$ and the matrix elements
$U_{IJ}$ are $\theta$-dependents and reduce
respectively to 1 and $\delta_{IJ}$ for $\theta = 0$.\\
Since the matrix $\Omega$ is invertible, the action can be simply
read off from the symplectic structure evident in the deformed
commutation relation (67):
\begin{eqnarray}
S = \sum_{IJ} \int dtdx (\partial_0\Phi_I)
(\Omega^{-1})_{IJ}(\partial_1\Phi_J) -
(\partial_1\Phi_I)\delta_{IJ}(\partial_1\Phi_J).
\end{eqnarray}
It is easy to check that the above action implies the equations of
motion (70-71). At this stage some remarks are in order. First,
notice that the action (74) is formally similar to one derived in
[31]
\begin{eqnarray}
\sum_{IJ} \int dtdx (\partial_0\Phi_I) K_{IJ}(\partial_1\Phi_J) -
(\partial_1\Phi_I)V_{IJ}(\partial_1\Phi_J)
\end{eqnarray}
to characterize the topological orders and to classify the
different hierarchies in abelian fractional Hall effect. In
equation (75), the matrix $K$ is symmetric and $V$ must be a
positive definite
matrix.\\
According to this formal similarity,  the action given by (74)
provides us with an useful bridge between the quantum theory of
noncommutative chiral fields and fractional quantum Hall effect.
Of course, since the matrix $\Omega$ is arbitrary, the action (74)
offers the possibility to cover different fractional Hall
hierarchies. As first simple illustration, we consider the matrix
$\Omega$ having the following form
\begin{eqnarray}
\Omega_{IJ}(\theta) = a_{IJ}(\theta)\delta_{IJ} + (I-J)(b_{IJ}(\theta) - b_{JI}(\theta))
\end{eqnarray}
where $a_{IJ}$  and $b_{IJ}$ are real parameters. In particular, for
$N=2$ and

$$a_{IJ} = -\exp(i\frac{\pi}{2}(I+J))  {\hskip 1cm} b_{IJ} =
\frac{\theta}{2}\epsilon_{IJ},$$ we obtain
\begin{eqnarray}
\Omega_{IJ} = (-)^{I+1}\delta_{IJ} + (I-J)\theta \epsilon_{IJ}.
\end{eqnarray}
This is exactly the metric considered in [19]. We will next study
the effects of the non-commutativity of chiral fields space in the
context of fractional quantum Hall effect when the matrix encoding
the deformation is given by (77). This particular choice
generates, as it is shown in the next section, a deviation from
the Laughlin hierarchy and the obtained model can be used to
describe others hierarchies like Jain one.

\section{Generalized fractional quantum Hall filling}
In this section, we assume the left and right components of a
quantum Hall droplets are coupled. The matrix encoding this
coupling is given by (77) where $I=1$ (resp. $I = 2$) is replaced
by $s=+$ (resp. $s=-$). Thus, the left and right chiral fields
satisfy the commutation relation
\begin{eqnarray}
[\Phi_s(x) , \Phi_{s'}(y)] = i s(\epsilon_{ss'} \theta -
\delta_{ss'})\varepsilon(x-y)
\end{eqnarray}
and  from the equation (74) the action reads
\begin{eqnarray}
S = \frac{1}{\bar{\theta}}\sum_s \int dxdt \bigg[\sum_{s'}
\partial_1\Phi_s[s\delta_{ss'} -
s\theta\epsilon_{ss'}]\partial_0\Phi_{s'} - \bar{\theta}
(\partial_1\Phi_s)^2\bigg]
\end{eqnarray}
where  $\bar{\theta} = \sqrt{1 + \theta^2}$. It is important to note
that this result can be achieved using the  following transformation
\begin{eqnarray}
\Phi_s(x) = \sqrt{\frac{\bar{\theta}+1}{2}}\bigg( \phi_s(x) -
\sqrt{\frac{\bar{\theta}-1}{\bar{\theta}+1}}
\epsilon_{ss'}\phi_{s'}(x)\bigg)
\end{eqnarray}
where the $\phi_s$'s stands for un-deformed chiral fields defined
in the previous section. Remark that the hamiltonian (68) remains
unchanged under (80)
\begin{eqnarray}
 H = \sum_s H_s = \sum_s \int dx : (\partial_1\Phi_s)^2 = \sum_s \int dx :
(\partial_1\phi_s)^2:
\end{eqnarray}
Using the relation (80) and equations (70) and (78), we expand the
chiral fields
\begin{eqnarray}
\Phi_{s}(x,t) = \beta^s_0 + \bar{\beta}^s_0(\bar{\theta}t+sx) + i
\sum_{n\neq 0}\frac{\beta^s_n}{n} e^{-in(\bar{\theta}t+sx)}
\end{eqnarray}
where the operators $\beta$ satisfy the relations
\begin{eqnarray}
[\beta^s_0,\bar{\beta}^{s'}_0] =i(\delta_{ss'}-\theta\epsilon_{ss'})
{\hskip 0.5cm} [\beta^s_n , \beta^{s'}_m] =n\delta_{ss'}\delta_{n+m}
+ \theta n\delta_{n-m}\epsilon_{ss'}{\hskip 0.5cm} others = 0
\end{eqnarray}
of two non-commuting copies of $U(1)$ Kac-Moody algebra. The
non-commutativity between right and left sectors  disappears for
$\theta = 0$ and the relations (83) give (64) as it is expected.
We note that a mapping can be established between the deformed
oscillations operators (83) and un-deformed ones (64). This is
\begin{eqnarray}
\beta_0^s = \sqrt{\frac{\bar{\theta}+1}{2}}\bigg( \alpha_0^s -
\sqrt{\frac{\bar{\theta}-1}{\bar{\theta}+1}}
\epsilon_{ss'}{\alpha}^{s'}_0\bigg)
\end{eqnarray}

\begin{eqnarray}
\bar{\beta}_0^s = \sqrt{\frac{\bar{\theta}+1}{2}}\bigg(
\bar{\alpha}_0^s - \sqrt{\frac{\bar{\theta}-1}{\bar{\theta}+1}}
\epsilon_{ss'}{\bar{\alpha}}^{s'}_0\bigg)
\end{eqnarray}

\begin{eqnarray}
\beta_n^s = \sqrt{\frac{\bar{\theta}+1}{2}}\bigg( \alpha_n^s -
\sqrt{\frac{\bar{\theta}-1}{\bar{\theta}+1}}
\epsilon_{ss'}{\alpha}^{s'}_n\bigg).
\end{eqnarray}
To compute the Hall filling factor associated to the edge
excitations described by the action (79), we define the operators
\begin{eqnarray}
{\cal O}_s(x)  = \eta :e^{\frac{i}{\sqrt{2}}\gamma \Phi_s(x)}:
\end{eqnarray}
where $\eta$ is constant. Using
$$ :e^A::e^B: = e^{<AB>}:e^Ae^B:,$$
it is easy to verify that the operators (87) satisfy the exchange
relation
\begin{eqnarray}
{\cal O}_s(x)  {\cal O}_{s'}(x')=
e^{i\gamma^2\Omega_{ss'}\varepsilon(x-x')}{\cal O}_{s'}(x') {\cal
O}_s(x).
\end{eqnarray}
Since ${\cal O}_s(x)$ is identified as an electron operator, it must satisfy
\begin{eqnarray}
{\cal O}_s(x)  {\cal O}_{s}(x')= -{\cal O}_{s}(x') {\cal O}_s(x).
\end{eqnarray}
Thus
$$\gamma^2 = (2m+1) {\hskip 2cm} m\in{\bf N}$$
is an odd integer. However, for $s \neq s'$ we have
\begin{eqnarray}
{\cal O}_s(x)  {\cal O}_{s'}(x')=
e^{-i(2m+1)\pi\theta\epsilon_{ss'}\varepsilon(x-x')}{\cal
O}_{s'}(x') {\cal O}_s(x).
\end{eqnarray}
The exchange relation (90) is a consequence of the
non-commutativity between left and right sectors. The non local
operators ${\cal O}_s(x)$ describe hard core objects ($({\cal
O}_s(x))^2 = 0$) which obey the standard (fermionic)
anticommutation relations.  In view of the exchange relation (90),
we shall refer to the objects ${\cal O}_s(x)$ as generalized
fermionic operators. Note also that the
relation (90) is similar to the exchange relation for anyonic particles. \\
Now we come the Green correlation functions which play an
important role since they give  the filling factor the quantum
Hall system described by the action (79). They can be easily
evaluated. Indeed, using the mapping (84-86) expressing the
deformed modes in term of the un-deformed ones and the
transformation (80), the correlation functions can be expressed in
terms of the usual Green functions
\begin{eqnarray}
 \langle \phi_s(x)^{\dagger} \phi_{s'}(0)\rangle \sim -s\delta_{ss'}\ln| x + st|.
\end{eqnarray}
Thus after some algebra and by rescaling $\gamma^2 \to
\frac{2}{\pi}\gamma^2$ we get
\begin{eqnarray}
\langle {\cal O}_s^{\dagger}(x) {\cal O}_s(0) \rangle \sim
\frac{1}{(x + \bar{\theta}t)^{(2m+1)\bar{\theta}}}.
\end{eqnarray}
The first thing we see that the electron propagator on the edge of
a fractional Hall droplet acquires a non trivial exponent
$(2m+1)\bar{\theta}$ not equal to one. This implies that the
electrons on the edge are strongly correlated. Using the
terminology of quantum Hall physics, we will call this type of an
electron state the generalized Luttinger liquid. For $\theta = 0$,
the exponent $(2m+1)$ give the Laughlin sequence associated with
the fractional filling factors $ \nu = 1/2m+1$. We would like to
emphasize that the exponent is the product of the two factors:
$(2m+1)$ which is directly linked to the statistics (89) and
$\bar{\theta}$ can be regarded as parameter characterizing some
topological interactions (responsible of anyonic-like relation
(90)) in the bulk of quantum Hall droplet. Thus, it is clear that
for edges excitations described by the action (79), the anomalous
exponent in (92) gives the following fractional filling factor
\begin{eqnarray}
\nu = \frac{1}{(2m+1)\bar{\theta}}.
\end{eqnarray}
 More
importantly, since there is no restriction on the deformation
matrix $\Omega$ in (74) and consequently on  the parameter
$\theta$ in (79), one can cover many others filling sequences
obtained from a theoretical point of view or observed
experimentally. Hence by an appropriate choice of the elements of
the matrix elements $\Omega$ one can reproduce the fractional
filling factors for other sequences like Jain ones [32]. Indeed,
setting
$$\bar{\theta} = 1 - \frac{1}{2m+1}\bigg(1 - \frac{1}{p}\bigg)$$
where $p \in {\bf N}- \{0\}$, one obtains the filling factors
$$\nu = \frac{p}{2m+p}$$
corresponding to stable fractional quantum Hall states on the Jain
sequences. Note that for $p=1$, we have $\bar{\theta} = 1$ (i.e.
$\theta = 0$) and the Jain sequences coincide with Laughlin
quantum Hall states. In this respect, the non-commutative chiral
boson action (79) describes a deviation ($\theta$-deformation)
 from the laughlin picture
and gives rise new fractional Hall states.

\section {\bf Left and right interacting fields and effective action}
In the previous sections, we started with a deformed commutation
rules to define the noncommutative chiral boson fields. In this
section, we consider two coupled chiral fields $\phi_+$ and
$\phi_-$ and we show that the interacting system is described by a
noncommutative action of type (79). It seems that the coupling
between the right and left sectors induces naturally the
non-commutativity in the fields space. We show here again that the
coupling
modifies the velocities of the chiral fields $\phi_+$ and $\phi_-$. \\
The uncoupled system is described by the  action
\begin{eqnarray}
S_0 =\int dtdx [ k_+(\partial_0\phi_+)(\partial_1\phi_+)-
(\partial_1\phi_+)^2 - k_-(\partial_0\phi_-)(\partial_1\phi_-) -
(\partial_1\phi_-)^2]
\end{eqnarray}
where  the real constants $ k_{\pm}$ are given by $k_{\pm} =\frac{1}{v_{\pm}}$ in terms of the fields velocities.
By adding a coupling term to take into account
the interaction between the fields $\phi_+$ and $\phi_-$
\begin{equation}
S_c = \int dtdx  k [(\partial_1\phi_+)(\partial_0\phi_{-}) +
(\partial_1\phi_-)(\partial_0\phi_{+})],
\end{equation}
the total action $S = S_0 + S_c$ can be expressed in the following
compact form:
\begin{equation}
S = \int dtdx
\sum_{ss'}(\partial_1\phi_s)\omega_{ss'}(\partial_0\phi_{s'}) -
\sum_s (\partial_1\phi_s)^2
\end{equation}
where
\begin{equation}
\omega_{ss'} = sk_s\delta_{ss'} + sk\epsilon_{ss'}.
\end{equation}
It is clear from the last two equations that the action (96) is
similar to one given by (79). The coupling constant play the role
of the non-commutativity parameter. The model described by the
action (96) is exactly solvable. In fact, by introducing the
fields
\begin{equation}
\phi_s = \sum_{s'}U_{ss'}\hat{\phi}_{s'}
\end{equation}
by mean of the matrix $U$ which diagonalizes the matrix $\omega$,
it is simply verified that the action (96) becomes
\begin{equation}
S = \int dtdx \sum_s [\lambda_s
(\partial_1\hat{\phi}_s)(\partial_0\hat{\phi}_s) ) -
(\partial_1\hat{\phi_s})^2]
\end{equation}
where
\begin{equation}
\lambda_{\pm} = \frac{1}{2}( k_+ - k_- ) \pm \frac{1}{2} \sqrt{4k^2
+ (k_+ + k_-)^2}
\end{equation}
are the the eigenvalues of the matrix $\omega$. Notice that under
the transformation $U$, we get a system of two uncoupled  chiral
fields travelling with different velocities $\lambda_+ \neq
\lambda_-$. Further, since
$$\lambda_+  \lambda_- = \det(\omega) < 0$$
the fields $\hat{\phi}_1$ and $\hat{\phi_2}$ are travelling in
opposites directions. Note also that the transformation $U$ leaves
the Hamiltonian
$$ H = \sum_s (\partial_1\phi_s)^2 = \sum_s (\partial_1\hat{\phi_s})^2$$
unchanged.  The new fields $\hat{\phi_s}$ satisfy the deformed commutations relations
\begin{equation}
[\hat{\phi}_{s} , \hat{\phi}_{s'}] = -i \Delta_{ss'}
\varepsilon(x-y)
\end{equation}
where
$$ \Delta_{ss'} = \sum_{s"} s"(U^{-1})_{ss"}(U^{-1})_{s's"},$$
and the matrix $U$ is given by
\begin{equation}
 U\,=\,
\pmatrix{\frac{k_- + \lambda_+}{\sqrt{k^2 +
 (k_-+\lambda_+)^2}} & \frac{k}{\sqrt{k^2 + (k_-+\lambda_-)^2}}
  \cr \frac{k}{\sqrt{k^2 + (k_-+\lambda_+)^2}}
 & \frac{k_-+\lambda_-}{\sqrt{k^2 + (k_-+\lambda_-)^2}} \cr}\,.
\end{equation}
in terms of the shifted velocities $\lambda_+$ and $\lambda_-$.
Subsequently, the system of the coupled chiral fields $\phi_+$ and
$\phi_-$ can be described by the action (96) where the fields
satisfy the canonical commutation relations (61) or alternatively
by the action (99) with chiral fields satisfying deformed
commutation rules (101). The two descriptions are equivalents. It
is important to notice that the coupling between the left and
right chiral fields induces a modification of their velocities
$$ k_+ \longrightarrow \lambda_+  {\hskip 2cm} k_- \longrightarrow
\lambda_- .$$ Finally, It is remarkable that for $k = 0$, we have
$U_{ss'} \to \delta_{ss'}$ and $\Delta_{ss'}\to s\delta_{ss'}$. In
this limit the commutations relations (101) reduce to un-deformed
ones (61). This agrees with the generalized deformation procedure
discussed in the subsection 5.2.

\section{Nonlinear chiral boson dispersion}
Recently there has been considerable interest in the generalization
of the chiral Luttinger liquid (CLL). This is essentially motivated
by experimental observations which argue a noticeably difference
with CLL prediction [34-35]. From a theoretical point of view, the
generalization is based on the interplay between the
electron-electron interaction and the confining potential at the
edge. This gives rise new additional low energy modes. In this
sense, in a recent works, it was shown analytically [33] and
numerically [36] that the chiral boson dispersion becomes nonlinear
when the non local nature of the electron-electron interaction is
incorporated in the CLL theory.\\
In this short section, we shall show  that the
theory of noncommutative fields gives rise the nonlinear dispersion
relation corroborating the results mentioned above. For this, let us
consider the fields
\begin{eqnarray}
\hat{\Phi}_s = \phi_s + \theta \sum_{s'}\epsilon_{ss'}\pi_{s'}
\end{eqnarray}
expressed in terms of the un-deformed ones. This transformation is
formally similar to one given by equation (21) or (49). The new
fields
 satisfy the deformed commutation rules
\begin{eqnarray}
[\hat{\Phi}_s(x) , \hat{\Phi_{s'}}(x') ] = -2i\theta
\epsilon_{ss'}\delta(x-x')-i\delta_{ss'}s(\varepsilon(x-x')+
\theta^2\delta '(x-x'))
\end{eqnarray}
which reduce the canonical ones in the limiting case $\theta = 0$. We
assume that the Hamiltonian takes the form
\begin{eqnarray}
H =\frac{1}{4\pi} \sum_s \int dx :(\partial_1\hat{\Phi}_s)^2:
\end{eqnarray}
in term of the new fields. Substituting (103) in (105), $H$ can be
cast in the following form
\begin{eqnarray}
H = \frac{1}{4\pi}\sum_s \int dx :[(\partial_1\phi_s)^2 + \theta^2
(\partial_1^2\phi_s)^2+
\theta\sum_{s'}\epsilon_{ss'}(\partial_1\phi_s)(\partial_1^2\phi_{s'})]:
\end{eqnarray}
in term of the un-deformed fields. Clearly edge dynamics described
by the Hamiltonian (106) with fields satisfying the canonical
commutation relations  is equivalent to the description provided
by $H$ Eq.(105) in terms of deformed fields obeying the
commutation rules (104). The effect of non-commutativity is now
apparent in the expression of the Hamiltonian. Using the mapping
(103) and the fields expansion (60), we obtain
\begin{eqnarray}
H = \frac{1}{2}\sum_s[(\bar{\alpha}^s_0)^2 +
2\sum_{n>0}(1+\theta^2n^2)\alpha^s_{-n}\alpha^s_n]
\end{eqnarray}
describing chiral bosons with nonlinear dispersion relation
\begin{eqnarray}
E_n \sim n+\theta^2n^3
\end{eqnarray}
which coincides with the result obtained in [33] in which the author incorporates the
electron-electron interaction in the
chiral Luttinger liquid theory. This  constitutes another illustration of
our interest for the quantum theory of noncommutative fields.
\section{Concluding remarks}

Two facets of quantum theory of noncommutative scalar fields are considered. \\
The first one, developed in the first part of this paper, concerns
the theory of noncommutative scalar fields $\varphi^1(x,t)$ and
$\varphi^2(x,t)$. We have shown that the non-commutativity at the
quantum level is deeply related to the deformation of the
symplectic structures
 of the classical phase space. The
quantization of the theory is performed thanks to a mapping which is called dressing
transformation. We stress that our approach is based
on the symplectic deformation and establish clearly that the noncommutative theory of
 fields originates from the deformation of the corresponding phase space.\\
The second facet related to noncommutative fields deals with the
non-commutativity between left and right sectors of a chiral boson
fields. We develop a general approach to introduce a theory of
noncommutative chiral fields. As a particular case, we recover the
model discussed in [19] which is shown to be useful in the context
of fractional quantum Hall effect. Indeed, the non-commutativity
generates a deviation from the Laughlin fractional filling factor
and gives, for some selected values of deformation parameter, the
factors associated with Jain sequence. Note that the action (74)
is formally similar to (75) which constitutes now the basis of our
understanding of the Hall hierarchies classification [31]. We
derived the action describing the coupling between the right and
left excitations living on the edges of a quantum Hall sample. It
is remarkable that the coupled system, described by the action
(96), is equivalent to one describing noncommutative chiral field
theory. The coupling induces a shift of the velocities of chiral
components. Finally, we show that the non-commutativity in the
fields space generate a nonlinear dispersion relation. This is in
agreement with recent analytical and numerical analysis [33,36].
To close this paper, we mention that due to the similarity between
the action (74) and (75), the quantum theory of noncommutative
fields constitutes another suitable approach to classify the Hall
hierarchies. The results obtained throughout this paper illustrate
the interest of the quantum theory of noncommutative fields,
especially in the context
 of fractional quantum Hall effect. This open the way for further
 investigations. Note also that the present study can be extended to the situation
 where the fields are self interacting which constitutes one of
 the main features of quantum field theory. We hope to report on these issues in a forthcoming work.

{\vskip 1cm}
 \noindent {\bf Acknowledgments}\\
MD would like to express his thanks to Max Planck Institute for
Physics of Complex Systems (MPI-PKS) where this work was done. The
authors are indebted to the referee for his constructive comment.

\end{document}